\documentclass[aps,prd,twocolumn,preprintnumbers,amsmath,amssymb,superscriptaddress,groupedaddress,floatfix,nofootinbib]{revtex4}

\usepackage[normalem]{ulem}
\usepackage[utf8]{inputenc}
\usepackage{epsfig,psfrag,graphics,verbatim}
\usepackage{graphicx}
\usepackage[T1]{fontenc}

\usepackage{dcolumn}
\usepackage{bm}
\usepackage{xcolor}
\usepackage{float}
\usepackage{adjustbox}
\usepackage{amsmath}
\usepackage{subcaption}
\usepackage{makecell}

\usepackage{varioref}
\usepackage{hyperref}
\usepackage{cleveref}

\date{March 2022} 

\begin{document}
\title{Testing $f(R)$ gravity models with quasar x-ray and UV fluxes}

\author{Matías Leizerovich}
\email{mleize@df.uba.ar}
\affiliation{Departamento de Física, Facultad de Ciencias Exactas y Naturales, Universidad de Buenos Aires \\
Av. Intendente Cantilo S/N 1428  Ciudad Autónoma de Buenos Aires, Argentina }
\affiliation{Consejo Nacional de Investigaciones Científicas y Técnicas (CONICET), Godoy Cruz 2290, 1425, Ciudad Autónoma de Buenos Aires, Argentina}
\author{Lucila Kraiselburd}
\affiliation{Consejo Nacional de Investigaciones Científicas y Técnicas (CONICET), Godoy Cruz 2290, 1425, Ciudad Autónoma de Buenos Aires, Argentina}
\affiliation{%
 Facultad de Ciencias Astronómicas y Geofísicas, Universidad Nacional de La Plata, Observatorio Astronómico, Paseo del Bosque,\\
B1900FWA La Plata, Argentina \\
}%
\author{Susana Landau}
\affiliation{Departamento de   Física, Facultad de Ciencias Exactas y Naturales, Universidad de Buenos Aires, \\
Av. Intendente Cantilo S/N, 1428,  Ciudad Autónoma de Buenos Aires, Argentina }
\affiliation{IFIBA - CONICET - UBA \\
Avenida Intendente Cantilo S/N, 1428, Ciudad Autónoma de Buenos Aires, Argentina}
\author{Claudia G. Scóccola}
\affiliation{Consejo Nacional de Investigaciones Científicas y Técnicas (CONICET), Godoy Cruz 2290, 1425, Ciudad Autónoma de Buenos Aires, Argentina}
\affiliation{%
 Facultad de Ciencias Astronómicas y Geofísicas, Universidad Nacional de La Plata, Observatorio Astronómico, Paseo del Bosque,\\
B1900FWA La Plata, Argentina \\
}%

\begin{abstract}
Recently, active galactic nuclei (AGN) have been proposed as "standardizable candles," thanks to an observed nonlinear relation between their x-ray and optical-ultraviolet (UV) luminosities, which provides an independent measurement of their distances. In this paper, we use these observables for the first time to estimate the parameters of $f (R)$ gravity models (specifically the Hu-Sawicki  and  the exponential models) together with the cosmological parameters. The importance of these types of modified gravity theories lies in the fact that they can explain the late time accelerated expansion of the Universe without the inclusion of a dark energy component.  We have also included other observable data to the analyses such as estimates of the Hubble parameter $H (z)$  from cosmic chronometers (CCs), the Pantheon Type Ia (SnIa) supernovae compilation, and baryon acoustic oscillations (BAO) measurements. The $1 \sigma$ inferred constraints using all datasets are $b \le 0.276$, $\Omega_m=0.304^{+0.010}_{-0.011}$ and $H_0=67.553^{+1.242}_{-0.936}$ for the Hu-Sawicki model, and $b= 0.785^{+0.409}_{-0.606}$ $\Omega_m=0.305^{+0.011}_{-0.010}$, and $H_0=68.348^{+0.959}_{-0.760}$ for the exponential one, but we stress that for both $f(R)$ models results within $2\sigma$ are consistent with the $\Lambda$ cold dark matter ($\Lambda$CDM) model. Our results show that the allowed space parameter is restricted when both AGN and BAO data are added to CC and SnIa data, with the BAO dataset being the most restrictive one. We can conclude that both the $\Lambda$CDM model and  small deviations from general relativity given by the $f(R)$ models studied in this paper, are allowed by the considered observational datasets.

\end{abstract}
\maketitle

\section{Introduction}

The late time accelerated expansion of the Universe is still one of the most intriguing conundrums that any successful cosmological model has to explain. 
In 1998, two international teams (Riess \textit{et al.} \cite{Riess_1998,Schmidt1998TheHS} and Perlmutter \textit{et al.} \cite{Perlmutter1998MeasurementsO})
showed independently observational evidence of this phenomena. Since then, great efforts have been made in order to explain the physical mechanism responsible for it. In the standard cosmological model [$\Lambda$ cold dark matter ($\Lambda$CDM)], a cosmological constant $\Lambda$ is added to the Einstein equations of general relativity:

\begin{equation}
R_{\mu\nu}-\frac{R}{2}g_{\mu\nu}+\Lambda g_{\mu\nu}=\kappa T_{\mu\nu}\text{,}\label{eq:Einstein_lambda}
\end{equation}
where $R_{\mu\nu}$ is the Riemann tensor, $R$ is the Ricci scalar, $g_{\mu\nu}$ is the metric, $\kappa =8 \pi G$ (for $c=1$) and $T_{\mu\nu}$ is the energy-momentum tensor. 
However, this proposal has several problems that have been discussed in the literature. For instance, the observational value of the cosmological constant $\Lambda$ does not match the value that is expected from theoretical estimations by 60-120 orders of magnitude \cite{Weinberg1989,Bousso2007,Carroll2001,SAHNI2000}.
In this context, alternative cosmological models have been considered to  provide an  explanation for the dynamics of the Universe's expansion.
These models can be classified into two families \cite{CLIFTON20121}: those which incorporate scalar fields with minimal coupling to gravity and matter (for example, quintessence or k-essence fields \cite{Joyce2016, 2013CQGra..30u4003T}) and  those which are based in alternative gravity theories. In the last group we find theories like Gauss-Bonnet, Horndeski and the so-called $f(R)$ theories \cite{PhysRevD.76.044027,Horndeski1974,Kobayashi2011,DeFelipe2010,CLIFTON20121}, among many others.
Another motivation for studying alternative cosmological models is the Hubble tension.
 Specifically, the value of the current Hubble parameter $H_0$ that has been obtained using cosmic microwave background (CMB) data and assuming a standard cosmological model \cite{Planckcosmo2018} is not in agreement with the one using model-independent observations, such as the luminosity from supernovae Ia \cite{Riess2019}.\footnote{There is also no agreement within the scientific community of the amount of this tension. While some authors claim that there is a  $4-\sigma$ tension \cite{Riess2019}, others claim lower amounts  or even no disagreements \cite{2021arXiv210511461M,2021ApJ...919...16F}.}

$f(R)$ theories \cite{DeFelipe2010}, despite originally being proposed by Starobinsky \cite{Starobinsky1980} in the 1980s to describe the inflation mechanism, have recently become relevant for explaining the late time accelerated expansion of the Universe. In these models, the Ricci scalar $R$ on the Einstein-Hilbert action is replaced by a scalar function of $R$. 
Although many $f(R)$ were proposed in the past, the vast majority of them have been ruled out by theoretical reasons such as antigravity regimes \cite{BAMBA2014136} or by experimental and observational constraints such as  local gravity tests   \cite{DeFelipe2010,TINO2020103772,Oikonomou2014} and solar system tests   \cite{Sotiriou2010,Faulkner2007,Capozziello2008,GUO2014,Chiba2007}. Two models that are still considered viable are the Hu-Sawicki  \cite{2007PhRvD..76f4004H} and the exponential ones \cite{PhysRevD.77.046009,Odintsov2017,PhysRevD.91.044019}.\\
Recently, Desmond and Ferreira \cite{2020PhRvD.102j4060D}, by using morphological indicators in galaxies to constrain the strength and range of the fifth force, have claimed that the Hu-Sawicki $f(R)$ model  can be ruled out. 
In their methodology, they use general relativity (GR)-based mock catalogs to which the effects of the $f(R)$ model are added. However, the results obtained by superimposing analytical expressions for the $f(R)$ effects to a $\Lambda${\rm CDM}  cosmology are different from those obtained from a modified-gravity-based simulation,  such as those presented in~\cite{2018MNRAS.480.5211N}.

In Nunes \textit{et al.} \cite{2017JCAP...01..005N} different $f(R)$ models (including the Hu-Sawicki and the exponential models) have been tested using cosmic chronometers (CCs), baryon acoustic oscillations (BAOs), joint light curves samples from supernovae Ia (SnIa) and astrophysical estimates of $H_0$. 
Also, Farugia \textit{et al.} \cite{Farrugia2021} have analyzed the same $f(R)$ models using several of the observational data mentioned above (but updated) plus redshift space distorsions (RSD) dataset and model-dependent CMB data.
In D'Agostino and Nunes \cite{PhysRevD.100.044041,PhysRevD.101.103505}, the Hu-Sawicki model has been tested with newer datasets such as gravitational waves and lensed quasars from the H0LICOW Collaboration.
 In Odintsov \textit{et al.} \cite{Odintsov2017} a change of variables to express Friedmann equations for the exponential model has been proposed while in Ref. \cite{ODINTSOV2021115377} a comparison between their numerical solution and the latest updates of the aforementioned observational data  has been made. In the present work, we constrain the Hu-Sawicki and the exponential models using a large set of cosmological observations, including, for the first time for these models, a recently released dataset of active galactic nuclei (AGN) compiled from Lusso \textit{et al.} \cite{LR2020} and taking the astrophysical parameters $\beta$, $\gamma$ and $\delta$ from Li \textit{et al.} \cite{Li2021}. This dataset together with SnIa and BAO data has recently been considered by Bargiacchi \textit{et al.} \cite{2021arXiv211102420B} to constrain the $\Lambda$CDM model as well as extensions of the latter and to discuss implications for nonflat cosmological models.

This paper is organized as follows: In Sec. II we briefly describe the main aspects of the $f(R)$ models in the cosmological context, we recall the  modified Friedmann equations and we present the $f(R)$ models that are analyzed in this paper.  In Sec. III, we describe the observational data that are used to test the predictions of the theoretical models. We also explain the statistical treatment that we have chosen for  the AGN data  which is based in the one proposed in Ref. \cite{Li2021}. In Sec. IV, we present the results of the statistical analyses. Comparison with similar works is discussed  in Sec. V while the conclusions are presented in Sec. VI. Each $f(R)$ model considered in this paper requires a specific change of variables to solve Friedmann equation. We describe the details of this procedure in the Appendix.

\section{Theoretical models}
\label{theory}
The ${\it f}(R)$ theories refer to a set of gravitational theories whose Lagrangian is given by a function of the Ricci scalar $R$, where each ${\it f} (R)$ defines a different model. Therefore, the Einstein-Hilbert action for these theories is
\begin{equation}
S = \frac{1}{2\kappa} \int d^{4}x \sqrt{-g} f(R)+S_{m}+S_{r},
\end{equation}
 where $S_m$ and $S_r$ represent the matter and radiation actions, respectively. 
The field equations are obtained by varying the action $S$ with respect to the metric $g_{\mu\nu}$ such that
\begin{equation}
R_{\mu\nu}f_{R}-\frac{1}{2}g_{\mu\nu}f\left(R\right)+\left(g_{\mu\nu}\Box-\nabla_{\mu}\nabla_{\nu}\right)f_{R}=\kappa T_{\mu\nu}\label{eq:Einstein-f(R)} ,
\end{equation}
 where ${\it f_R}=\frac{d {\it f}}{d R}$, $\Box$ is the d'Alembertian operator, $\nabla_{\mu}$ is the covariant derivative, and $T_{\mu\nu}$ is the energy-momentum tensor.

In this work we assume a spatially flat Friedmann-Lema{\^i}tre-Robertson-Walker (FLRW) cosmology so the metric  is  given by
\begin{equation}
ds^{2}=-dt^{2}+a^{2}\left(t\right)\left(dr^{2}+r^{2}d\Omega^{^{2}}\right),  
\end{equation}
where $a(t)$ is the scale factor of the Universe and $H=\dot{a}/a$ is the Hubble parameter (the dot represents the derivatives with respect to the cosmic time). Then, the Ricci scalar can be written as
\begin{align}
R & =6\left(2H^{2}+\dot{H}\right)\text{.}\label{eq:Ricci(H)}
\end{align}

Considering the energy-momentum tensor of a perfect fluid   $T^{\mu}_{\nu}={\rm diag}(-\rho,P,P,P)$  (where $\rho=\rho_m+\rho_r$ and $P=P_m+P_r$), the field equations \eqref{eq:Einstein-f(R)} become
\begin{subequations}\label{eq:Friedmanns}
\begin{align}
\label{eq: Friedmann_1}
-3H^{2} & = -\frac{1}{f_{R}}\left[\kappa\rho+\frac{Rf_{R}-f}{2
}-3H\dot{R}f_{RR}\right]\\
\label{eq: Friedmann_2}
-2\dot{H} & = \frac{1}{f_{R}}\left[\kappa\left(\rho+P\right)+f_{RRR}\dot{R}^{2}+\left(\ddot{R}-H\dot{R}\right)f_{RR}\right],
\end{align}
\end{subequations}
where $f_{RR}$ and $f_{RRR}$ are the second and third derivative with respect to $R$, respectively. It has been shown that the latter equations  can be expressed as a set of first order equations, which results in a more stable system from the numerical point of view \cite{2016PhRvD..93h4016D,Odintsov2017}. There are numerous proposals in this regard. In this article we assume the change of variables proposed in Ref. \cite{Odintsov2017} for the exponential model and the one used by de la Cruz-Dombriz \textit{et al.} \cite{2016PhRvD..93h4016D} for the Hu-Sawicki model. Both settings are described in the Appendix.

The continuity equations of matter and radiation for a flat FLRW metric can be expressed as
\begin{equation}\label{eqrho}
\dot{\rho}+3H\left(\rho+P\right)=0.
\end{equation}
At redshifts between $0\leq z \leq 10^4$, considering pressureless (nonrelativistic) matter and radiation (relativistic particles), the solution is $\rho=\rho_m^0 a^{-3}+\rho_r^0 a^{-4}$.\\

 Viable $f(R)$ models must fulfill some theoretical constraints such as having a positive gravitational constant, stable cosmological perturbations, and avoiding ghost states, among many others \cite{2007PhRvD..76f4004H,2010LRR....13....3D,2017JCAP...01..005N}. Therefore,
to elude instabilities when curvature becomes too large at high densities\footnote{Other requirements that are usually asked to ensure stability in scenarios with large curvatures are  $\lim_{R \to 0} {\it f} (R) -R = 0$ and $\lim_{R \to \infty} {\it f} (R) -R$ = constant.},
it is necessary that
\begin{equation}
f_R>0 \,\,\,\, \text{and} \,\,\,\,  f_{RR}>0, \,\,\,\,\, \ensuremath{\text{for }\,R\geq R_{0}},
\end{equation}
where $R_0$ is the  current value of the Ricci scalar.
Moreover, as discussed previously,  a successful cosmological model must  provide an explanation for the late accelerated expansion of the Universe. For this, it is required that   ${\it f} ( R )\rightarrow R-2\Lambda$ when $R\geq R_{0}$, where $\Lambda$ is an effective cosmological constant. 
On the other hand,  bounds from local tests of gravity such as solar system and equivalence principle tests require that a viable $f(R)$ model shows a "chameleonlike" mechanism \cite{Brax2008,Hui2009,Negrelli2020}. 
Lastly, the stability of a late time de Sitter solution must be guaranteed. Consequently,the following condition has to be fulfilled
\begin{equation}
0<\frac{Rf_{RR}}{f_{R}}\left(r\right)<1\,\,\,\text{at}\,\,\, r=-\frac{Rf_{R}}{f}=-2 .
\end{equation}
Accounting for all these restrictions, the viable models can be expressed as follows,
\begin{equation}
f(R)=R-2\Lambda y(R,b)\text{,}\label{eq:cv_generico}
\end{equation}
with $y(R,b)$ a function that quantifies the deviation from GR and $b$ the distortion parameter that quantifies the effect of that deviation.

As a consequence of the  restrictions described above, the behavior of these  ${\it f}(R)$ models tends asymptotically to the one of the $\Lambda$CDM at large redshifts ($z\geq 10^4$), when the curvature $R$ also becomes large \cite{2007PhRvD..76f4004H,2010LRR....13....3D,Odintsov2017,2007JETPL..86..157S}. 
However, the late time evolution of these theories differs from $\Lambda$CDM. 
Hence, if 
${\Omega_i}=\kappa\rho_{i}^0 /3{H_0}^2$ 
is the current critical density, where $H_0$ and $\rho_{i}^0$ refer to  the current  values of the Hubble parameter and density $\rho_{i}$,
these quantities ($\Omega_{i}$ and $H_0$)  defined in ${\it f} (R)$ models, will be different from the same quantities defined in the $\Lambda$CDM model. 
Still, all these quantities are related through the physical matter density \cite{2007PhRvD..76f4004H},
\begin{equation}\label{relationmodels}
{\Omega_m}{H_0}^2={\Omega^{\Lambda {\rm CDM}}_m}\left({H_0}^{\Lambda {\rm CDM}}\right)^2=\frac{\kappa}{3}\rho_{m}^0.
\end{equation}
Besides, for the $\Lambda$CDM model it holds that
\begin{equation}\label{LCDMrelation}
{\Omega^{\Lambda {\rm CDM}}_m}+{\Omega^{\Lambda {\rm CDM}}_{\Lambda}}=1 ,
\end{equation}
 where ${\Omega^{\Lambda {\rm CDM}}_{\Lambda}}= \Lambda /3\left( H_0^{\Lambda {\rm CDM}}\right)^2$. It should be noted that the systems of differential equations that we use in this paper are written in terms of $\Omega_m^{\Lambda {\rm CDM}}$ and $H_0^{\Lambda {\rm CDM}}$ while the results of the statistical analyses will be reported in terms of the corresponding parameters defined in $f(R)$ models.
Equations \eqref{relationmodels} and \eqref{LCDMrelation} will be  useful to establish the initial conditions of the Friedmann equations. For this, the main assumption is that at high redshift the behavior of $H(z)$ in the $\Lambda$CDM and  $f(R)$ models is the same. Since the observational data used in this work are at  redshifts $z <8$, the radiation terms can be neglected.

Next, we present the two ${\it f} (R)$ models analyzed in this paper:
\begin{enumerate}
    \item {The exponential $f(R)$ model was proposed by Cognola \textit{et al.} \cite{PhysRevD.77.046009} and further discussed in  \cite{Odintsov2017,PhysRevD.80.123528,PhysRevD.91.044019}, among many others. In this model, the proposed $f(R) $ function can be expressed as:
\begin{equation}
f(R)=R - 2 \Lambda (1-e^{-\frac{R}{\Lambda b}}) ,
\end{equation}
where $b$  and $\Lambda$ are the free parameters of the model.}
\item{The currently known as Hu \& Sawicki model was developed by these authors in 2007~\cite{2007PhRvD..76f4004H}. The proposed $f(R)$ function can be expressed as:
\begin{equation}
f(R)=R-\frac{c_1R_{HS}(R/R_{HS})^n}{c_2(R/R_{HS})^n+1} ,
\end{equation}
where $c_1$, $c_2$, $R_{HS}$ and $n$ represent the free parameters of the model. 
It is possible to rewrite the above expression as the one proposed in Eq. $\eqref{eq:cv_generico}$,
\begin{equation}\label{HSf}
f(R)=R-2\Lambda\Big[1-\frac{1}{1+\left(\frac{R}{\Lambda b}\right)^n}\Big]   
\end{equation}
with $\Lambda=c_1R_{HS} / 2c_2$ and $b=2c_2^{1-1/n} /c_1$. It is easy to see that when $b\to 0$, the model reduces to a $\Lambda$CDM cosmology; ${\it f}(R)\to R-2\Lambda$. In this work, we will restrict ourselves to analyze only the case when $n = 1$.}
\end{enumerate}
Finally, as mentioned above, the system of equations that we choose to  solve to obtain $H(z)$ in each model as well as the initial conditions and the details of dealing with numerical instabilities will be described later in the Appendix.

\section{Observational Data}
\label{Observational Data}
In this section, we present the datasets that we use to determine the values of the ${\it f}(R)$ parameters that best fit the different cosmological observations.

\subsection{Cosmic chronometers}

The CC is a method developed by Simon \textit{et al.} \cite{simon05} that allows one to determine the Hubble parameter $H(z)$ from the study of the differential age evolution of old elliptical passive-evolving\footnote{Passive evolving means that there is no star formation or interaction with other galaxies.} galaxies  that formed at the same time but are separated by a small redshift interval. The method relies on computing the Hubble factor $H(z)$ from the following expression:
\begin{equation}
H(z)=\frac{-1}{1+z}\frac{dz}{dt},
\end{equation}
where $dz/dt$ can be calculated from the ratio $\Delta z/\Delta t$  and  $\Delta$ refers to the difference  between the two galaxies whose properties have been described above.

The galaxies chosen for this method were formed early in the Universe, at high redshift ($z > 2-3$), with large mass ($\mathcal{M}_{\rm stars}>10^{11}\mathcal{M}_{\odot}$), and their stellar production has been inactive since then.
Hence, by observing the same type of 
galaxies at late cosmic time, stellar age evolution can be used as a clock synchronized with cosmic time evolution. On the other hand, $dz$ is determined by spectroscopic surveys with high precision. The goodness of this method lies in the fact that the measurement of relative ages $dt$ eliminates the systematic effects present in the determination of absolute ages. Furthermore,  $dt$ is independent of the cosmological model since it only depends on atomic physics and not on the integrated distance along the line of sight (redshift).

For this work, we use the most precise available estimates of  $H(z)$, which are summarized in Table \ref{tab:CC}.

\bgroup
\def\arraystretch{1.3}
\begin{table}
\resizebox{0.3\textwidth}{!}{
\begin{tabular}{| c | c | c | c |}
    \hline
    $z$ & $H(z) \ (\mathrm{km}\,\mathrm{s}^{-1}\,\mathrm{Mpc}^{-1}) $ & Reference \\
    \hline
     0.09 & 69  $\pm$  12   & \\
    0.17 & 83  $\pm$  8    &  \\
    0.27 & 77  $\pm$  14   & \\
    0.4  & 95 $\pm$ 17     & \\
    0.9  &117 $\pm$ 23     &~\cite{simon05}  \\
    1.3  &168 $\pm$ 17     & \\
    1.43 &177 $\pm$ 18     & \\
    1.53 &140 $\pm$ 14     & \\ 
    1.75 &202 $\pm$ 40     & \\
    \hline
    0.48 & 97  $\pm$  62   &~\cite{stern10} \\
    0.88 & 90  $\pm$  40   & \\
    \hline  
    0.1791 &  75  $\pm$  4 & \\ 
    0.1993 & 75  $\pm$  5  & \\
    0.3519 & 83  $\pm$  14 & \\
    0.5929 & 104  $\pm$  13 &~\cite{moresco12} \\
    0.6797 & 92  $\pm$  8  & \\
    0.7812 & 105  $\pm$  12 & \\
    0.8754 & 125  $\pm$  17 & \\
    1.037  & 154  $\pm$  20 & \\
    \hline   
    0.07   & 69  $\pm$  19.6   &  \\
    0.12   & 68.6  $\pm$  26.2 &~\cite{zhang14} \\  
    0.2    & 72.9  $\pm$  29.6 & \\
    0.28   & 88.8  $\pm$  36.6 & \\
    \hline 
    1.363  & 160  $\pm$  33.6  &~\cite{moresco15}  \\
    1.965  & 186.5  $\pm$  50.4 & \\ 
    \hline 
    0.3802 & 83  $\pm$  13.5   &  \\
    0.4004 & 77  $\pm$  10.2   & \\
    0.4247 & 87.1  $\pm$  11.2 &~\cite{CC2}\\
    0.4497 & 92.8  $\pm$  12.9 & \\
    0.4783 & 80.9  $\pm$  9    & \\
\hline 
\end{tabular}}
\caption{$H(z)$ estimates from the cosmic chronometers. Each column stands for the redshift of the measurement, the $H(z)$ mean value (and its standard deviation)  and  reference, respectively.}
\label{tab:CC}
\end{table}

\subsection{Supernovae type Ia}

Type Ia supernovae are one of the most luminous events in the Universe, and are considered as standard candles due to the homogeneity of both its spectra and light curves.
As we will explain below, the distance modulus $\mu$ can be determined from the SnIa data, and alternatively, it can also be described as,
\begin{equation}
\mu=25+5 \log_{10}(d_L(z)),
\label{distmod}
\end{equation}
where $d_L$ the luminosity distance
\begin{equation}
d_L(z)=(1+z)\int_0^z\frac{dz'}{H(z')}.
\label{distlum}
\end{equation}
Since the previous expression shows how this last magnitude depends on both the redshift $z$ and the cosmological model [via $H(z)$], it is possible to compare the distance modulus predicted by the theories with the estimates from observations.

In this case, we are considering  1048 SnIa at redshifts between $0.01 < z < 2.3$ from the Pantheon compilation~\cite{2018ApJ...859..101S}. For this compilation, the observed distance modulus estimator is expressed as,
\begin{equation}
\mu=m_B-M+\alpha x_1+\beta c +\Delta_M+\Delta_B,
\label{mu}
\end{equation}
with  $m_B$ being an overall flux normalization, $x_1$ the deviation from the average light-curve shape, and $c$ the mean SnIa B-V color index.\footnote{Parameters $m_B$, $x_1$ and $c$ are determined from a fit between a model of the spectral sequence SnIa and the photometric data (for details, see \cite{2018ApJ...859..101S}).} Meanwhile, $M$ refers to the absolute B-band magnitude of a fiducial SnIa with $x_1 = 0$ and $c = 0$, and $\Delta_B$ refers to a distance correction based on predicted biases from simulations. Coefficients $\alpha$ and $\beta$ define the relations between luminosity and stretch and  between luminosity and color, respectively. 

On the other hand, $\Delta_M$ represent a distance correction based on the mass of the SnIa's  host galaxy. For this SnIa compilation, it is obtained from
\begin{equation}
\Delta_M=\gamma \times[1+e^{(-(m-m_{\rm step})/\tau)}]^{-1} ,
\end{equation}
where $m_{\rm step}$ is a mass step for the split, $\gamma$ is a relative offset in luminosity, and $m$ is the mass of the host galaxy. Parameter $\tau$ symbolizes an exponential transition term in a Fermi function that defines the relative probability of masses to be on one side or the other of the split. Both $m_{\rm step}$ and $\tau$ are derived from different host galaxies samples (for details, see~\cite{2018ApJ...859..101S}). Finally, coefficients $\alpha$, $\beta$, $M$, and $\gamma$ are the so-called nuisance parameters of the SnIa.

These parameters are usually determined through a statistical analysis with supernovae data  where a $\Lambda$CDM model is assumed. In particular Scolnic \textit{et al.} obtain  for the Pantheon sample \cite{2018ApJ...859..101S} the following values $\alpha=0.0154 \pm 0.006$, $\beta=3.02 \pm 0.06$, and $\gamma=0.053 \pm 0.009$. To verify these values, we have assumed the Hu-Sawicki model and performed a statistical analysis with the same dataset allowing both the nuisance  and  the model parameters to vary.\footnote{Given the strong degeneracies between the parameters when only SnIa data are used, we have considered a fixed value for $H_0$ (we have analysed two cases: one with $H_0=$67.4 km $\rm{s^{-1}}$ $\rm{Mpc^{-1}}$ \cite{Planckcosmo2018} and another one with $H_0=$ 73.5 km $\rm{s^{-1}}$ $\rm{Mpc^{-1}}$ \cite{Riess2018}.)}
Our estimated nuisance parameters are  consistent with those computed by the Pantheon compilation within $1 \sigma$. This agreement has been also obtained in  a similar analysis carried out assuming another alternative theory of gravity \cite{2020JCAP...07..015N}, and in \cite{2018ApJ...859..101S} where extensions of the $\Lambda$CDM models where assumed. All those mentioned analyses confirm that the value of the nuisance parameters are independent of the cosmological model.  Therefore, in all statistical analyses reported in Sec. \ref{results} we fix the nuisance parameters to the values published by the Pantheon compilation.

\subsection{Baryon acoustic oscillations}
\label{bao}

Before the recombination epoch, photons and electrons were coupled through Thomson scattering, generating sound waves in the primordial plasma. Once the temperature of the Universe has dropped sufficiently as for neutral hydrogen to form, matter and radiation decouples, and the acoustic oscillations are frozen, leaving an imprint both in the cosmic microwave background and in the distribution of matter at large scales.
The maximum distance that the acoustic wave could travel in the plasma before decoupling defines a characteristic scale, named the sound horizon at the drag epoch $r_d$. Hence, BAOs provide a standard ruler to measure cosmological distances. Several tracers of the underlying matter density field provide different probes to measure distances at different redshifts. 

 The BAOs signal along the line of sight directly constrains the Hubble constant $H(z)$ at different redshifts. When measured in a redshift shell, it constrains the angular diameter distance $D_A(z)$,
\begin{equation}
    D_A(z) = 
    \frac{1}{(1+z)} \int_0^z \frac{dz'}{H(z')}.
\end{equation}
To separate
 $D_A(z)$ and $H(z)$, BAO should be measured in the anisotropic 2D correlation function, for which extremely large volumes are necessary. If this is not the case, a combination of both quantities can be measured as

\begin{equation}
D_V(z) = \left[ (1+z)^2 D_A^2(z) \frac{z}{H(z)}  \right]^{1/3}.
\end{equation}

Currently, there are many precise measurements of BAOs obtained using different observational probes. In general, a fiducial cosmology is needed in order to measure the BAO scale from the clustering of galaxies, or any other tracer of the matter density field. It is known that the standard BAO analysis gives model-independent results, and that it can be used to perform cosmological parameter inference to constrain exotic models. In particular, Bernal et al~\cite{Bernal2020} have demonstrated the robustness of the standard BAO analysis when studying models whose extensions to the $\Lambda$CDM model may introduce contributions not captured by the template used. They have found no significant bias in the BAO analysis for the exotic models they studied.
The distance constraints presented in Table~\ref{tab:BAO_data} include information about $r_d^{fid}$, which is the sound horizon at  the drag epoch computed for the fiducial cosmology. 

\bgroup
\def\arraystretch{1.5}
\begin{table}
\resizebox{0.48\textwidth}{!}{
\begin{tabular}{| c | c | c | c |}
\hline 
 $z_{\rm eff}$ &  Value & Observable  & Reference  \\ 
\Xhline{2.5\arrayrulewidth}
$0.15$	& 	$4.473 \pm 0.159$   &  $D_V/r_d$  &	\cite{SDSS_DR7}\\
\hline
$0.44$	& 	$11.548 \pm 0.559$  &  $D_V/r_d$  &	\\
$0.6$	& 	$14.946 \pm 0.680 $  &  $D_V/r_d$  &	\cite{WiggleZ}\\
$0.73$	& 	$16.931 \pm 0.579$   &  $D_V/r_d$  &	\\
\hline
$1.52$	& 	$26.005 \pm 0.995$   &  $D_V/r_d$  &	\cite{SDSSIVquasars}\\
\Xhline{2.5\arrayrulewidth}
$0.81$	& 	$10.75 \pm 0.43$    &  $D_A/r_d$  &~\cite{DES_Y1}\\
\Xhline{2.5\arrayrulewidth}
$0.38$	& 	$10.272 \pm 0.135 \pm 0.074$   &  $D_M/r_d$    & \\
$0.51$	& 	$13.378 \pm 0.156 \pm 0.095$ &  $D_M/r_d$   &~\cite{BOSS}\\
$0.61$	& 	$ 15.449 \pm 0.189 \pm 0.108$    &  $D_M/r_d$   & \\
\hline
$0.698$	& 	$ 17.65 \pm 0.3$    &  $D_M/r_d$   & \cite{SDSSIVLRGanisotropicCF}\\
\hline
$1.48$	& 	$  30.21 \pm 0.79$    &  $D_M/r_d$   & \cite{SDSSIVQSOanisotropicPS} \\
\hline
$2.3$	& 	$37.77 \pm 2.13$   &  $D_M/r_d$  &		\cite{SDSSIIILaforests}\\
\hline
$2.4$	& 	$36.6 \pm 1.2$  &  $D_M/r_d$  &	\cite{Laforestsquasarscross}\\
\Xhline{2.5\arrayrulewidth}
$0.698$	& 	$ 19.77 \pm 0.47$    &  $D_H/r_d$   & \cite{SDSSIVLRGanisotropicCF}\\
\hline
$1.48$	& 	$  13.23 \pm 0.47$    &  $D_H/r_d$   & \cite{SDSSIVQSOanisotropicPS} \\
\hline
$2.3$	& 	$9.07 \pm 0.31 $  &  $D_H/r_d$  &	\cite{SDSSIIILaforests}\\
\hline
$2.4$	& 	$8.94 \pm 0.22$   &  $D_H/r_d$  &	\cite{Laforestsquasarscross}\\
\Xhline{2.5\arrayrulewidth}
$0.38$	& 	$12044.07 \pm 251.226 \pm 133.002 $  &  $H r_d$ [km/s]  & \\
$0.51$	& 	$ 13374.09 \pm 251.226 \pm 147.78$    &  $H r_d$  [km/s]    & ~\cite{BOSS}\\
$0.61$	& 	$ 14378.994 \pm 266.004 \pm 162.558 $    &  $H r_d$ [km/s]   & \\
\hline
\end{tabular}}
\caption{Distance constraints from BAO measurements of different observational probes. The table shows the redshift of the measurement, the mean value and standard deviation of the observable, the observable that is measured in each case and the corresponding reference.}
\label{tab:BAO_data}
\end{table}

Here we describe the observations used in this work.  
In  Ross \textit{et al.}~\cite{SDSS_DR7}, the main spectroscopic sample of Sloan Digital Survey data release 7 (SDSS-DR7) galaxies is used to compute the large-scale correlation function at $z_{\rm eff}= 0.15$. The nonlinearities at the BAO scale are alleviated using a reconstruction method.
The first year data release of the Dark Energy Survey~\cite{DES_Y1} measured the angular diameter distance $D_A/r_d$ at $z_{\rm eff}=0.81$, from the projected two point correlation function of a sample of $1.3\times 10^{6}$ galaxies with photometric redshifts, in an area of 1336 deg$^2$. 
The final galaxy clustering data release of the Baryon Oscillation Spectroscopic Survey~\cite{BOSS}, provides measurements of the comoving angular diameter distance $D_M/r_d$ [related with the physical angular diameter distance by $D_M(z) = (1+z) D_A(z)$] and Hubble parameter $H r_d$ from the BAO method after applying a reconstruction method, for three partially overlapping redshift slices centered at effective redshifts 0.38, 0.51, and 0.61. 
Measurements of $D_V/r_d$ at effective redshifts of  0.44, 0.6, and 0.73 are provided by the WiggleZ Dark Energy Survey~\cite{WiggleZ}.
With a sample of 147,000 quasars from the extended Baryon Oscillation Spectroscopic Survey (eBOSS) \cite{SDSSIVquasars} distributed over 2044 square degrees with redshifts $0.8 < z < 2.2$, a measurement of $D_V/r_d$ at $z_{\rm eff}= 1.52$ is provided.
The BAO can be also determined from the flux-transmission correlations in Ly$\alpha$ forests in the spectra of 157,783 quasars in the redshift range $2.1 < z < 3.5$ from the SDSS-DR12 \cite{SDSSIIILaforests}. Measurements of $D_M/r_d$ and the Hubble distance  $D_H/r_d$ [defined as $D_H = c/H(z)$] at $z_{\rm eff}=2.33$ are provided.
From the cross-correlation of quasars with the Ly$\alpha$-forest flux transmission of the final data release of the SDSS-III~\cite{Laforestsquasarscross}, a measurement of $D_M/r_d$ and $D_H/r_d$ at $z_{\rm eff}=2.4$ can be obtained.
From the anisotropic power spectrum of the final quasar sample of the SDSS-IV eBOSS survey~\cite{SDSSIVQSOanisotropicPS}, measurements for  $D_M/r_d$ and $D_H/r_d$ at $z_{\rm eff}=1.48$ are obtained. 
The analysis in the configuration space of the anisotropic clustering of the final sample of luminous red galaxies from the SDSS-IV eBOSS survey~\cite{SDSSIVLRGanisotropicCF} gives constraints on $D_M/r_d$ and $D_H/r_d$ at $z_{\rm eff}=0.698$.

\subsection{Quasar x-ray and UV fluxes}

Quasars are among the most luminous sources in the Universe. Besides, they are observable at very high redshift and therefore they are regarded as promising cosmological probes.  
In the last years, the observed relation between the ultraviolet  and x-ray
emission in quasars has been used to develop a new method to convert quasars into standardizable candles \cite{RL2015,RL2019,LR2020}. In this work, we will use the recent compilation provided by Risaliti and co-workers \cite{LR2020} of x-ray and UV flux measurements of
2421 quasars quasi-stellar object (QSOs/AGN) which span the redshift range $0.009 \leq z \leq 7.5413$ to test the cosmological models based in alternative theories of gravity described in Sec. \ref{theory}. The relation between the quasar  UV and x-ray luminosities can be described by the following equation:
\begin{equation}
    \log L_{\rm X} = \gamma \log L_{\rm UV} + \beta_1 ,
    \label{luminosities}
\end{equation}
where $L_{\rm X}$ and $L_{\rm UV}$ refer to the rest-frame monochromatic luminosities at 2 keV and 2500 \r{A} respectively. The constants $\gamma$ and $\beta_1$ are  determined with observational data and should be independent of redshift in order to assure the robustness of the method \cite{RL2015,RL2019,LR2020}.  It was pointed out in \cite{LR2020} that there is a strong correlation between the parameters involved in the quasar luminosity relation and cosmological distances, therefore, in order to test cosmological models, luminosity distances obtained from quasar fluxes should be cross-calibrated previously using, for example, data from type Ia supernovae.
In this work, we use the calibration method proposed by Li \textit{et al.} \cite{Li2021} which uses a Gaussian process regression to reconstruct the expansion history of the Universe from the latest type Ia supernova observations.\footnote{Within the Gaussian process, a theoretical model is assumed but it has been discussed in \cite{Li2021} that the results are independent from this choice.} Next, we will  briefly describe  how this method, which is almost model-independent,  is implemented. Equation \eqref{luminosities} can be expressed in terms of the UV and x-ray fluxes as follows:
\begin{equation}
    \log F_{\rm X} =\gamma  \log F_{\rm UV} + 2 (\gamma-1) \log (d_L H_0) + \beta ,
    \label{fluxes}
\end{equation}
where $d_L$ refers to the luminosity distance and $\beta=\beta_1 + (\gamma -1 ) \log 4\pi - 2 (\gamma -1) \log H_0$\footnote{It should be stressed that the parameter $\beta$ defined here is different from the one in Refs \cite{RL2019,LR2020}.}. From Eq. \eqref{fluxes}, the quantity  $ \log F_X^{\rm SN}$ can be defined and computed, using quasar measurements of $F_{\rm UV}$, while  the quantity $d_L H_0$ is obtained from a Gaussian Process regression  method with the latest SnIa data \cite{Li2021}.
Furthermore, the following likelihood is assumed:
\begin{equation}
\ln {\cal L}= - \dfrac{1}{2} \sum_i \left(\frac{\log \left( F_X\left(\gamma,\beta\right)\right)_i^{\rm SN}-\log \left(F_X\right)_i^{\rm QSO} }{s_i^2} \right) + \ln s_i^2 ,
\end{equation}
where $s_i^2=\sigma^2_{\log F_X}+\gamma^2 \sigma^2_{\log F_{UV}}+\delta^2$ and $\delta$ is an intrinsic dispersion that is introduced to alleviate the Eddington bias \cite{RL2019,LR2020}. In such way, considering  the X ray fluxes from quasar data ($\log F_X^{\rm QSO}$), Li \textit{et al.} \cite{Li2021} obtained $\gamma$, $\beta$, and $\delta$ in a model-independent way.  Their results are consistent within $1 \sigma$ with the ones obtained in \cite{LR2020} using Eq.~\eqref{fluxes}. Moreover, the independence of the $L_{\rm X} - L_{\rm UV}$ relation with redshift has been analyzed in  previous works \cite{RL2015,RL2019,LR2020}. 
In such way, we will test the cosmological models defined in Sec. \ref{theory}, using the following likelihood and assuming the values  of $\gamma$ and $\beta$  obtained in \cite{Li2021} ($\gamma=0.648 \pm 0.007$, $\beta=7.730\pm 0.244$)\footnote{It should be noted that the considered value of $\gamma$ is in agreement with the one obtained in Ref. \cite{2021arXiv211102420B} (G. Bargiachi private communication) where also the AGN, SnIa and BAO datasets are used and extensions of the $\Lambda$CDM cosmological models are considered. However, in the statistical analyses of Ref. \cite{2021arXiv211102420B} $\gamma$ is free to vary together with the cosmological parameters. Regarding the  parameter  $\beta$, there is not a fair comparison to be made since  the parameter $\beta$ in Ref \cite{2021arXiv211102420B} refers to the parameter $\beta_1$ in Eq. \eqref{luminosities} and it is necessary to fix the  value of $H_0$ to relate both parameters.}:

\begin{equation}
\ln {\cal L} =   - \dfrac{1}{2} \sum_i \frac{\left[ \log(d_L H_0\left({\bf\theta}\right) )^{\rm TH}_i - \log (d_l H_0)^{\rm QSO}_i\right]^2}{\sigma^2_{\log(d_L H_0)}} ,
\end{equation}
where {\bf $ \theta$}$=\left(\Omega_m^{\Lambda {\rm CDM}}, H_0^{\Lambda {\rm CDM}},b\right)$, $\log(d_L H_0\left({\bf\theta}\right) )^{\rm TH}$ refers to the theoretical prediction of the luminosity distance,  $\log (d_L H_0)^{\rm QSO}_i$ is calculated from Eq.~\eqref{fluxes} and 

\begin{eqnarray}
\sigma^2_{\log(d_L H_0)} &=& \frac{\sigma^2_{F_{\rm X}}+\gamma^2 \sigma^2_{F_{\rm UV}}+\sigma^2_\beta}{\left[2(\gamma - 1)\right]^2 }\nonumber\\
&+&\frac{\left(\beta +\log F_{\rm UV}-\log F_X\right)^2\sigma^2_\gamma}{\left[2(\gamma - 1)^2\right]^2} .
\end{eqnarray}

On the other hand, it has been argued recently that some of the subsamples of the dataset provided in \cite{LR2020} are not standardizable and have model and/or redshift dependence \cite{KR2021,KR2021_2,Luongo2021}. First of all, the analysis used to reach such a conclusion does not include any previous cross-calibration with supernovae data. It should also be noted that the most important differences in the values of $\gamma$ and $\beta$  obtained in these works are for models with different geometries, i.e, flat and nonflat models. Moreover, the present work is restricted to flat $f(R)$ models.  Furthermore, these analyses intend to constrain $\Omega_m$ and $H_0$ at the same time and it is well known that this cannot be done when using only data with information about the luminosity distances.

\section{Results}
\label{results}

In this section we present the results of our statistical analysis for both models; Hu-Sawicki with $n=1$ (HS) and the exponential $f(R)$ (EM).  As we have described in Sec. \ref{theory} the free parameters of these models  are: the distortion parameter $b$, the mass density $\Omega_m$, and the Hubble parameter $H_0$. 
To do the statistical analysis, we use a Markov chain Monte Carlo method and  the observational data described in Sec. \ref{Observational Data}. In cases where the SnIa observational data are used, $M_{abs}$ is also set as a free parameter. The priors used in this work are $H_0 \in [60,80]$, $\Omega_m \in [0.01,0.4]$, $M_{abs} \in [-22,-18]$,
and $b \in [0,5]$ for EM while for HS $b \in [0,2]$.
To perform the numerical integration and the statistical analyses we developed our own \textit{Python} code which uses \textit{Scipy}\cite{2020SciPy-NMeth} and \textit{Emcee}\cite{emcee} Python libraries and is publicly available in a \textit{Github} repository\cite{matiascode}.

 Table \ref{tab:results} and Fig. \ref{fig:f1} show the results for the two $f(R)$ models detailed in Sec. \ref{theory}  and the datasets described in Sec. \ref{Observational Data}. Furthermore, we include the results for the $\Lambda$CDM model for comparison. 

\begin{table*}[bt]
\begin{tabular}{llccccccc}
\cline{1-7}
                                                &               & $M_{abs}$                     & $\Omega_{m}$ & $b$ & $H_{0}$ &  &  &  \\ \cline{1-7}
\textbf{$\Lambda$CDM}                           & CC+SnIa         &                                                                                                                                       
                                                $-19.379^{+0.056(0.109)}_{-0.053(0.104)}$ &  
                                                $0.301^{+0.019(0.041)}_{-0.022(0.038)}$ & $-$ & $69.034^{+1.687(3.629)}_{-2.000(3.648)}$ &  &  &  \\
                                                & CC+SnIa+AGN     & $-19.407^{+0.058(0.103)}_{-0.049(0.107)}$ & $0.327^{+0.016(0.034)}_{-0.019(0.036)}$ & $-$ & $67.813^{+1.728(3.399)}_{-1.775(3.465)}$        &  &  &  \\
                                                & CC+SnIa+BAO     & $-19.395^{+0.024(0.051)}_{-0.025(0.049)}$ & $0.297^{+0.010(0.021)}_{-0.011(0.021)}$ & $-$ & $68.564^{+0.689(1.411)}_{-0.722(1.428)}$        &  &  &  \\
                                                & CC+SnIa+BAO+AGN & $-19.384^{+0.025(0.048)}_{-0.025(0.051)}$ & $0.306^{+0.010(0.020)}_{-0.011(0.019)}$ & $-$ & $68.786^{+0.729(1.469)}_{-0.729(1.404)}$        &  &  &  \\
                                                &               &                               &              &     &         &  &  &  \\
                                                \cline{1-7} 
\textbf{HS}                                     & CC+SnIa           &         
                                                $-19.374^{+0.054(0.103)}_{-0.051(0.105)}$ &     
                                                $0.269^{+0.036(0.059)}_{-0.028(0.062)}$ & 
                                                $\le 0.623(1.348)$ & $69.004^{+1.746(3.482)}_{-1.837(3.602)}$ &  &  &  \\
                                                & CC+SnIa+AGN     & $-19.409^{+0.052(0.105)}_{-0.052(0.100)}$ & $0.322^{+0.018(0.037)}_{-0.018(0.035)}$ & 
                                                $\le0.150(0.398)$ &
                                                $67.622^{+1.656(3.344)}_{-1.751(3.403)}$ &  &  &  \\
                                                & CC+SnIa+BAO     & 
                                                $-19.436^{+0.037(0.066)}_{-0.032(0.071)}$ & $0.292^{+0.012(0.022)}_{-0.011(0.022)}$ & 
                                                $0.294^{+0.084(0.400)}_{-0.269(0.294)}$ &
                                                $66.950^{+1.389(2.247)}_{-1.041(2.436)}$ &  &  &  \\
                                                & CC+SnIa+BAO+AGN & $-19.414^{+0.034(0.060)}_{-0.029(0.064)}$ & $0.304^{+0.010(0.020)}_{-0.011(0.021)}$ & 
                                                $\le 0.276(0.583)$ &
                                                $67.553^{+1.242(2.029)}_{-0.936(2.255)}$ &  &  &  \\
                                                &               &                               &              &     &         &  &  &  \\
                                                \cline{1-7}  
\textbf{EM}                                    & CC+SnIa        & 
                                                $-19.376^{+0.055(0.108)}_{-0.054(0.109)}$ &
                                                $0.293^{+0.025(0.049)}_{-0.022(0.051)}$ &
                                                $\le 1.102(2.015)$ &
                                                $68.998^{+1.880(3.705)}_{-1.850(3.621)}$ &  &  &  \\
                                                & CC+SnIa+AGN     &
                                                $-19.403^{+0.055(0.104)}_{-0.052(0.105)}$ &
                                                $0.324^{+0.019(0.038)}_{-0.019(0.037)}$ &
                                                $\le 0.749(1.272)$ &
                                                $67.903^{+1.699(3.410)}_{-1.789(3.434)}$ &  &  & 
                                                \\                                                & CC+SnIa+BAO     &
                                                $-19.405^{+0.031(0.055)}_{-0.025(0.056)}$ &
                                                $0.298^{+0.011(0.022)}_{-0.011(0.021)}$ &
                                                $\le 1.155(1.928)$ & 
                                                $68.011^{+1.136(1.833)}_{-0.777(2.079)}$ &  &  & 
                                                \\                                                & CC+SnIa+BAO+AGN     &
                                                $-19.393^{+0.028(0.053)}_{-0.026(0.055)}$ &
                                                $0.305^{+0.011(0.020)}_{-0.010(0.021)}$ &
                                                $0.785^{+0.409(0.760)}_{-0.606(0.785)}$ &
                                                $68.348^{+0.959(1.704)}_{-0.760(1.771)}$ &  &  & 
                                                \\
                                                &               &                               &              &     &         &  &  &  \\ 
\end{tabular}

\caption{Results from statistical analysis using data from CCs, luminosity distances reported by Pantheon collaboration (SnIa), AGN UV and x-ray luminosities, and several datasets from cosmological distances of BAOs. For each parameter, we present the mean value and the 68\% (95\%) confidence levels, or the upper limits obtained.}
\label{tab:results}
\end{table*}

\begin{figure*}
    \centering 
        \begin{subfigure}{0.49\textwidth}
            \includegraphics[width=\textwidth]{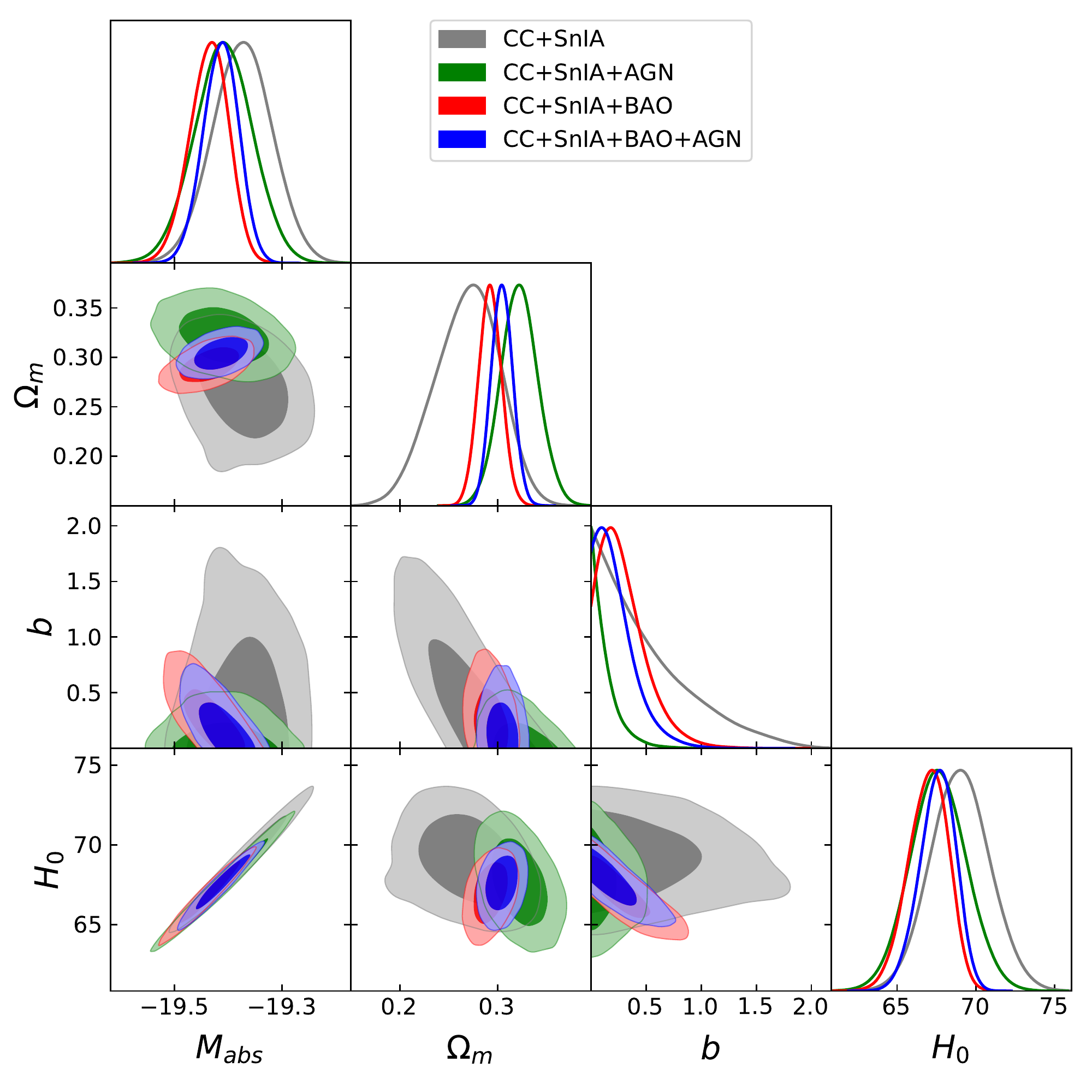}
        \end{subfigure}
       \hfill
        \begin{subfigure}{0.49\textwidth}
            \includegraphics[width=\textwidth]{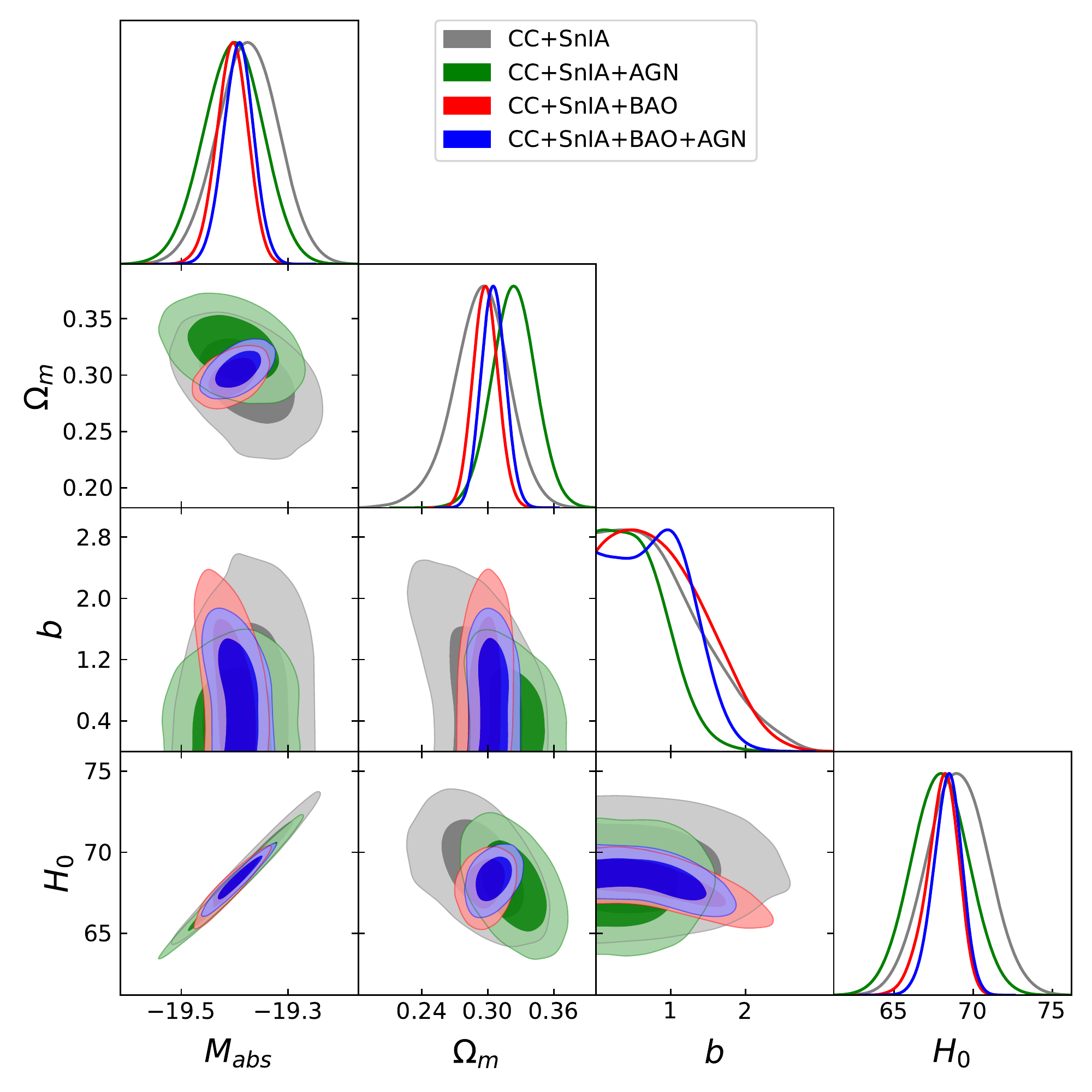}
        \end{subfigure}
    \caption{Results of the statistical analysis for the $f(R)$ Hu-Sawicki model (left) and the exponential Model (right). The darker and brighter regions correspond to 65\% and 95\% confidence regions, respectively. The plots in the diagonal show the posterior probability density for each of the free parameters of the model.}
    
    \label{fig:f1}
\end{figure*}
\subsection{The Hu-Sawicki model}

We emphasise that when the AGN or BAO data are added to the CC+SnIa analysis, the allowed parameter space is considerably reduced. We note that the BAO dataset is much more restrictive than AGN. Nevertheless, the constraining power of AGN is clearly seen.  Besides, the AGN data shift the fitted value of $\Omega_m$ to larger values (this fact has been already mentioned in \cite{Li2021} for the $\Lambda$CDM model) and the estimated $H_0$  to lower values.
We also notice that the shift on $\Omega_m$  (to larger values) and  $H_{0}$ (to lower values) is much more pronounced for AGN than for BAO.

Regarding the relation between $b$ and $H_0$, we mention that BAO data constrain the parameter space in such a way that there is a negative correlation between them. Besides, it follows from Fig. \ref{fig:f1} that  $\Omega_m$ and $b$ show degeneracies   when  CC and SnIa are considered and also where the AGN data are added to the latter.  We also remark that BAO reduces the allowed region of $\Omega_m$ considerably. 
Moreover, we  note that the correlation between $\Omega_m$  and  $H_{0}$ changes sign when BAO data are used, independent of whether the AGN data are used or not.

Lastly, for all datasets detailed in Table \ref{tab:results}, the $b$ values presented are consistent with zero ($\Lambda$CDM prediction) within 1$\sigma$, except for the case where CC, SnIa and BAO data were used together, in which the concordance is given at 2$\sigma$. The rest of the estimated free parameters  are in agreement with those obtained for the $\Lambda$CDM model.

\subsection{The exponential model}
 We note that the behavior of this model is very similar to the Hu-Sawicki one regarding the constraining power of the BAO and AGN datasets. In fact, the constraints on $\Omega_m$, $b$, and $H_0$ are considerably reduced when either dataset is included in the analysis, BAO being the most restrictive one. Also, we note that  the values of $\Omega_m$ and $H_0$ are also shifted when the AGN dataset is added in the same way described  previously for the  HS model.

As regards the correlation between parameters, there is no clear relation between $b$ and $H_0$ and the same is observed for the case of $b$ and $\Omega_m$. Conversely, the inclusion of BAO data makes the correlation between $\Omega_m$ and $H_0$ to change sign, the same effect we have already discussed for the HS model. 

On the other hand, we point out that the obtained intervals for the distortion parameter $b$ are larger than the ones of the Hu-Sawicki model. This is expected since it is necessary a bigger change on $b$ (in EM) to notice a difference with the $\Lambda$CDM model predictions. Furthermore, the constraints on $H_0$ and $\Omega_m$ are in agreement with those obtained for $\Lambda$CDM model for all statistical analyses carried out in this paper. Besides, the estimated $b$ constraints are consistent at 1$\sigma$ with the $\Lambda$CDM model ($b = 0$), except for the case where the CC+SnIa+BAO+AGN data were used; for the latter the consistency is within 2$\sigma$.  


Figure \ref{fig:f2} shows that the allowed parameter space for $\Omega_m$ and $H_0$ is enlarged with respect of the $\Lambda$CDM case and also the sign of the correlation changes when either the HS or the exponential model are considered. 
Finally, we also note that the parameter spaces obtained for the Hu-Sawicki and the exponential $f(R)$ models are compatible at 1$\sigma$ in all the studied cases. 

\begin{figure*}
    \centering 
    \includegraphics[width=0.4\textwidth]{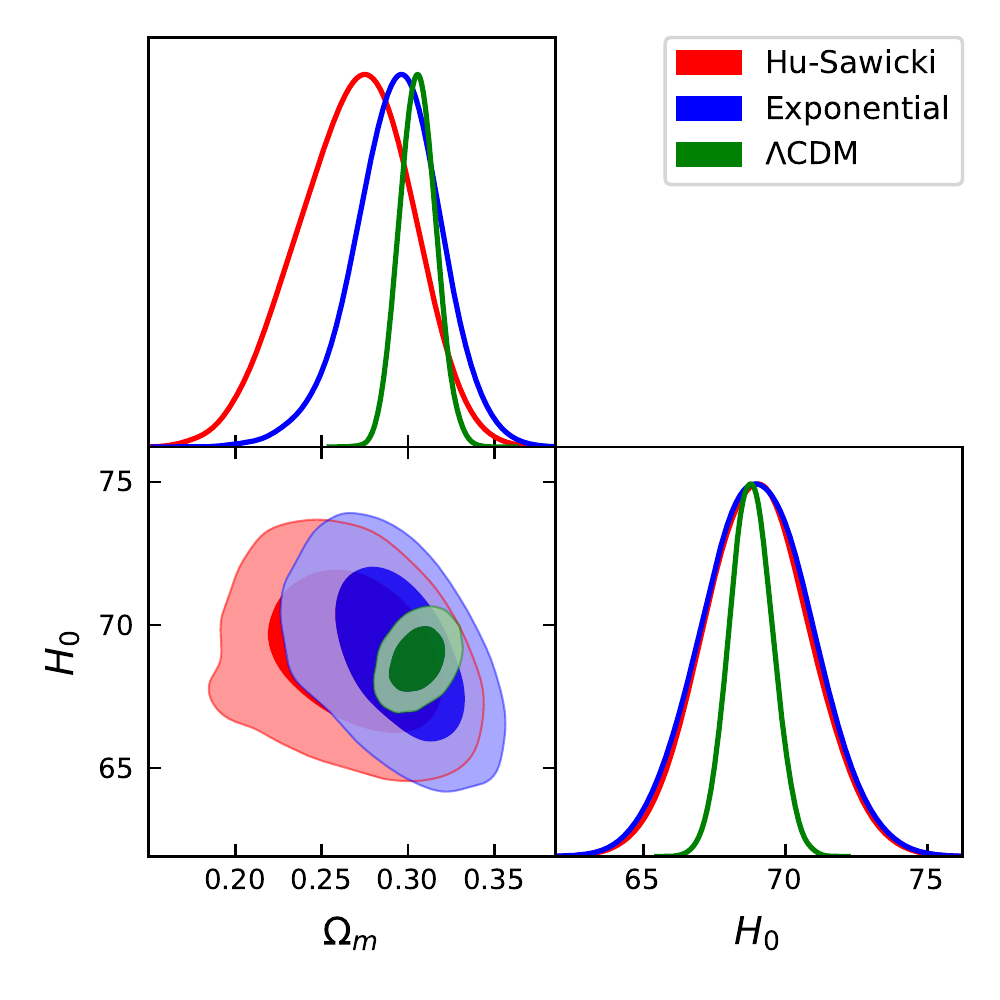}
    \caption{Results for the matter density $\Omega_{m}$ and the Hubble parameter $H_{0}$ using all the datasets (CC+SnIa+BAO+AGN). The plots show the 68\% and 95\% confidence region together with the posterior probability density for each parameter obtained  for the two $f(R)$ models considered in this paper and the $\Lambda \rm{CDM}$ model.}
    
    \label{fig:f2}
\end{figure*}

\section{Discussion}

Here we compare our results shown in the previous section with others that have already been published by other authors for the same $f(R)$ models using the same and/or similar datasets (\cite{PhysRevD.101.103505,HOLICOW2019,Farrugia2021} for HS, and \cite{ODINTSOV2021115377,Farrugia2021} for EM). We show in Fig.\ref{fig:intervals_HS} a comparison of our results with the ones obtained by other authors for the Hu-Sawicki model and the same is done in Fig.\ref{fig:intervals_EXP} for the exponential model.

Our parameter estimates for the Hu-Sawicki model using CC+SnIa data are 1$\sigma$ consistent with the ones published in \cite{PhysRevD.101.103505} for the same data compilations. The $b$ values reported in there are slightly smaller at 1$\sigma$ and smaller at 2$\sigma$ than ours. These differences are due to the fact that in that work, the authors only use the series expansion proposed by Basilakos \textit{et al.} \cite{PhysRevD.87.123529} to obtain an expression for $H (z)$\footnote{Private communication with R. C. Nunes.}, while we use the combination of methods explained in Sec. \ref{sec:HS} of the Appendix. That series expansion only allows them to explore  a small range of  $b$ values ($b<1$) which does not deviate much from the $\Lambda$CDM prediction; this does not happen in our analysis where the parameter space to be examined is much larger. Furthermore, in that article another statistical analysis is performed incorporating  data from six systems of strongly lensed quasars analyzed by the H0LICOW Collaboration \cite{HOLICOW2019} to the data mentioned before (CC+SnIa). Comparing the results of this analysis with our own, it is noticed that (i) the ranges of $H_0$ are in agreement within 2$\sigma$ except for our study of CC + SnIa + BAO and CC + SnIa + BAO + AGN; (ii) the $\Omega_m$ intervals are consistent at 1$\sigma$ except for our CC + SnIa + AGN analysis, where they are consistent at 2$\sigma$; and (iii) all the $b$ ranges are compatible at 1$\sigma$. Another interesting result to compare with is the one published by Farugia \textit{et al.} \cite{Farrugia2021}. Although their results are in agreement with 
ours with 1$\sigma$, their estimated range of $b$ values is very small (of the order  $10 ^{ -4}$). They use the same data compilations as we do for CC and SnIa but our BAO dataset is different, plus they add data from RSD and CMB. It should be noted that the CMB data used in \cite{Farrugia2021} refer to the acoustic scale $l_A$, the shift parameter $R$, and the current baryon density $\omega_b = \Omega_b h^2$. However, these observables are obtained through a statistical analysis where a $\Lambda$CDM model is assumed.
Therefore, in our opinion, it is not correct to use these data to constrain alternative cosmological models.
\begin{figure*}
    \centering 
    \includegraphics[width=1\textwidth]{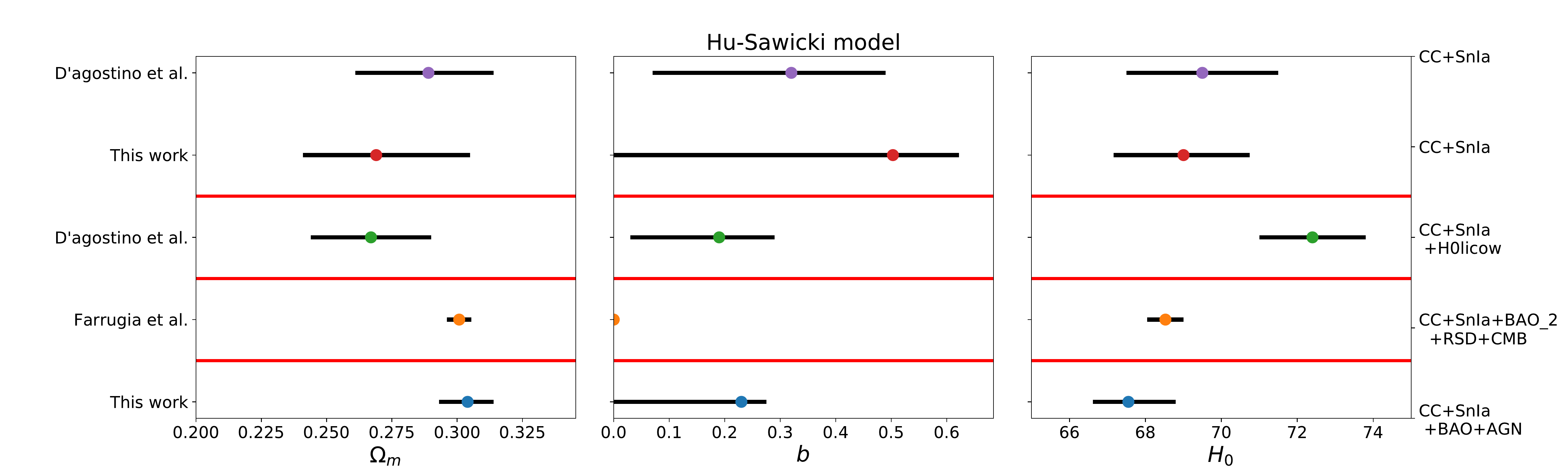}
    \caption{Constraints on the free parameters of the Hu-Sawicki model. Comparison between the 1$\sigma$ confidence intervals  obtained in this work and the ones reported by other authors.}
    \label{fig:intervals_HS}
\end{figure*}

\begin{figure*}
    \centering 
    \includegraphics[width=\textwidth]{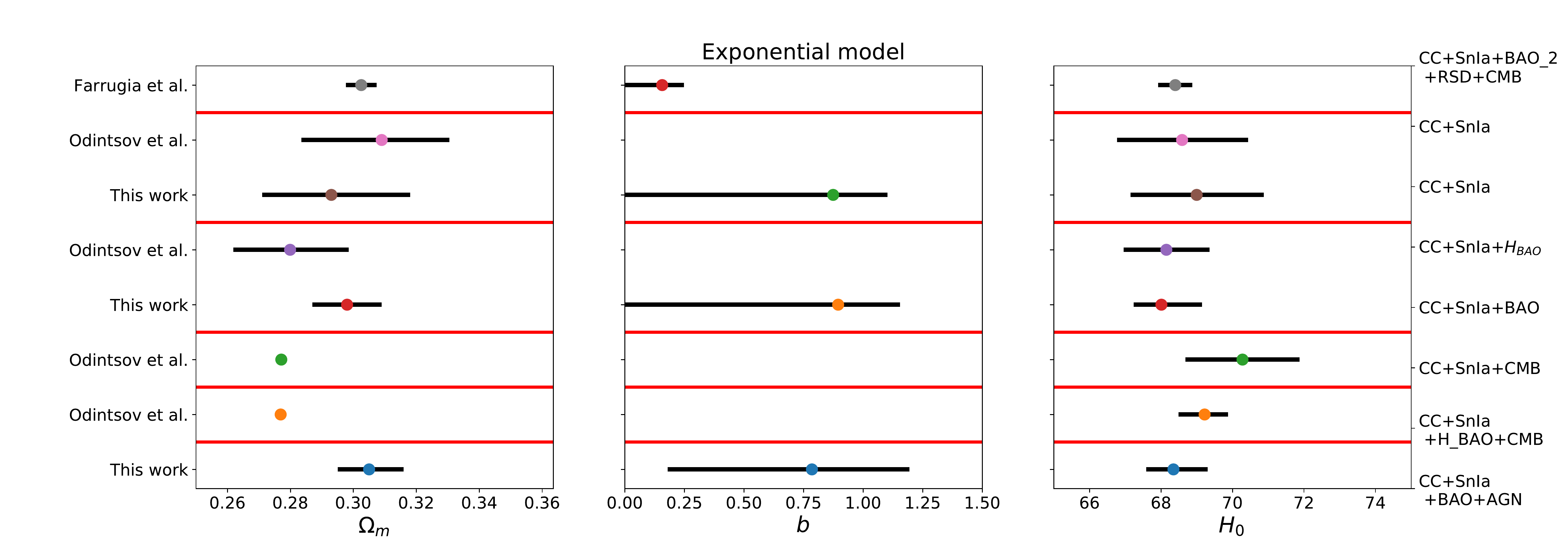}
    \caption{Constraints on the free parameters of the exponential model. Comparison between the 1$\sigma$ confidence intervals  obtained in this work and the ones reported  by other authors.}
    \label{fig:intervals_EXP}
\end{figure*}

 On the other hand, the estimates we have obtained for the parameters of the exponential $f(R)$ model using CC and SnIa data are consistent at 1$\sigma$ with the values of $\Omega_{m}$ and $H_{0}$ reported in \cite{ODINTSOV2021115377} for the same dataset. However, in that paper the $b$ interval is not reported, but it is for an associate quantity $\beta=2/b$. In order to compare it with our predictions, we tried to construct the posterior distribution for $\beta$ based on our distribution for $b$. Since the results are located near $b=0$, the distribution for $\beta$ tends to infinity  on the ranges of interest (as it is noticed in that article), so it cannot be sampled correctly. These authors also perform statistical tests using data from $H_{\rm BAO}$ (a BAO dataset different than ours) and CMB, which both further restrict the parameter space. Their estimates using CC+SnIa+$H_{\rm BAO}$ are compatible with ours (for CC+SnIa+BAO dataset) at 1$\sigma$, while their predictions using CC+SnIa+$H_{\rm BAO}$+CMB are consistent with ours (using CC+SnIa+BAO+AGN) within 1$\sigma$ only for the $H_0$ intervals, since the CMB data greatly reduce the $\Omega_m$ interval. Finally, in article \cite{Farrugia2021} a statistical analysis is also performed for the exponential model using the  CC+SnIa+BAO$_2$+RSD+CMB data (BAO$_2$ is a BAO dataset different than ours)  whose results are consistent within 1$\sigma$ with ours (using CC+SnIa+BAO+AGN) but their parameter intervals are narrower than ours. It should not be overlooked that the CMB data used in both papers \cite{Farrugia2021, ODINTSOV2021115377} are biased as explained above. 
Finally, from all the statistical analyses that have been performed in this paper, it is noted that for the  models studied here, the estimated $H_0$ parameters are consistent with the latest result reported by the Planck Collaboration \cite{Planckcosmo2018} within 1$\sigma$ but not with the ones published by Riess \textit{et al.} (\cite{Riess2018} and \cite{Riess2019}).
Besides, the obtained $\Omega_m$ confidence intervals are consistent with the ones obtained by the Planck Collaboration \cite{Planckcosmo2018} within 1$\sigma$ except for our CC+SnIa and CC+SnIa+BAO+AGN analyses with the Hu-Sawicki model where the agreement is within 2$\sigma$. 

\section{Conclusions}
\label{conclusions}
\par In this article we have analyzed two $f (R)$ models (HS and EM) in a cosmological context. For this, we have solved the corresponding Friedmann equations and we have performed statistical analyses considering recent datasets from SnIa, BAO, AGN and CC in order to constrain the free parameters of the models. The originality of this work lies in the use of AGN (not previously used for these particular theories) as standard candles to put bounds to the proposed models and the inclusion of the latest BAO data from the eBOSS Collaboration (2020). Furthermore, we have previously verified the consistency between the SnIa nuisance parameters published by the Pantheon Collaboration assuming a $\Lambda$CDM cosmological model and those estimated from the $f(R)$ models studied here.

Our results show that, although AGN narrow down the parameter space of cosmological models more than the SnIa and CC data, the baryon acoustic oscillation data continue to be the most restrictive ones. 
On the other hand, all our estimates for the different combinations of datasets are in accordance within 2$\sigma$ with the values reported by other authors for the same cosmological models but with different datasets. Moreover,  we  have  found  that  the $H_0$ estimates are consistent with the value reported by Planck Collaboration. The 1$\sigma$ obtained constraints when using the CC+SnIa+BAO+AGN dataset for the Hu-Sawicki model are $b \le 0.276$, $\Omega_m=0.304^{+0.010}_{-0.011}$ and $H_0=67.553^{+1.242}_{-0.936}$; and for the exponential model, $b= 0.785^{+0.409}_{-0.606}$, $\Omega_m=0.305^{+0.011}_{-0.010}$ and $H_0=68.348^{+0.959}_{-0.760}$. We stress that  results within 2$\sigma$ are in agreement with the $\Lambda$CDM model.

In summary, we have analyzed the Hu-Sawicki and the exponential $f(R)$ predictions with different and new datasets. Moreover, although the $b$ estimates are in agreement with the $\Lambda$CDM prediction at 2$\sigma$, the allowed region of the parameter space leads us to conclude that both HS and exponential $f(R)$ models are not yet ruled out by current data to explain the late time accelerated expansion of the Universe.

\section{Acknowledgments}
The authors would like to thank G.S. Sharov, R. Nunes, G. Bargiacchi, X. Li, S. Kandhai, H. Desmond, M. Salgado, B. Li, S. P\'erez Bergliaffa, E. Colg\'{a}in, and L. Perivolaropoulos  for their helpful comments.

The authors are supported by the National Agency for the Promotion of Science and Technology (ANPCYT) of Argentina Grant No. PICT-2016-0081, CONICET Grant No. PIP 11220200100729CO, and Grants No. G140, No. G157, and No. G175 from UNLP.
\par 

\appendix
\section{Solving the Friedmann Equations}
\label{EqFriedmannS}
In general, the Friedmann equations \eqref{eq:Friedmanns} are not easy to solve. In fact, it is a usual procedure to resolve them numerically. For this reason, it is desirable to improve the system stability and to speed up the computation time by choosing an appropriate parametrization for each model. In the following, we provide details of the numerical integration in each case including the initial conditions and the way of dealing with numerical instabilities.

\subsection{The exponential model}
For the exponential model it is very useful to express the Friedmann equations in terms of a new set of  variables as follows \cite{Odintsov2017}:
%
\begin{subequations}\label{eq:motion}
\begin{align}
\label{eq:motion1}
\frac{{\rm d} H}{{\rm d} x} & =\frac{R}{6H}-2H\\
\label{eq:motion2}
\frac{{\rm d} R}{{\rm d} x} & =\frac{1}{f_{RR}}\left(\frac{\kappa \rho}{3H^2}-f_{R}+\frac{Rf_R-f}{6H^2}\right)\\
\label{eq:motion3}
\frac{{\rm d} \rho}{{\rm d} x} & =-3(\rho+P).
\end{align}
\end{subequations}
Here $x ={\rm log} a=-{\rm log} (z+1)$  is the number of \textit{e}-folds, with $a(t_0)=1$ at the present time $t_0$. Using the following dimensionless change of variables,
\begin{equation}\label{varER}
E=\frac{H}{{H_0}^{\Lambda {\rm CDM}}},\,\,\,\,\,\,\,\,\,\,\,\, \mathcal{R}=\frac{R}{2\Lambda},
\end{equation}
the field equations are expressed in terms of the parameters ${\Omega^{\Lambda {\rm CDM}}_m}$, ${\Omega^{\Lambda {\rm CDM}}_{\Lambda}}$ and ${H_0}^{\Lambda {\rm CDM}}$ as
\begin{subequations}
\begin{align}
\begin{split}
\label{expE}
\frac{{\rm d}E}{{\rm d}x} & ={\Omega^{\Lambda {\rm CDM}}_{\Lambda}}\frac{\mathcal{R}}{E}-2E,
\end{split}\\
\begin{split}
\frac{{\rm d}\mathcal{R}}{{\rm d}x}&=\frac{2\Lambda}{f_{\mathcal{R}\mathcal{R}}}\Big[ \Omega^{\Lambda {\rm CDM}}_{m}\frac{a^{-3}+X^{\Lambda {\rm CDM}}a^{-4}}{E^2}\\
&-\frac{f_{\mathcal{R}}}{2\Lambda}+\frac{\mathcal{R}f_{\mathcal{R}}-f}{6\left(H_0^{\Lambda {\rm CDM}}\right)^2E^2}\Big], 
\label{expR}
\end{split}
\end{align}
\end{subequations}
%
where $X^{\Lambda {\rm CDM}}=\Omega^{\Lambda{\rm
CDM}}_{r}/\Omega^{\Lambda {\rm CDM}}_{m}$, and $f_{\mathcal{R}}$ and $f_{\mathcal{R}\mathcal{R}}$ are the first and second derivative with respect to $\mathcal{R}$.
This system of equations is solved numerically by establishing appropriate initial conditions. 

It has been already discussed that there are two situations in which the behavior of the model tends asymptotically to the $\Lambda$CDM solution: (i) high redshifts (large curvature) and (ii) $b \rightarrow 0$. Therefore, to perform the numerical integration  we can assume initial conditions that match the $\Lambda$CDM model at a redshift $z_i$ [$x_i=-\log(z_i+1)$], i.e.,
\begin{subequations}
\begin{align}
\label{condiniexpE}
E^2(x_i) & =\Omega^{\Lambda {\rm CDM}}_{m} \left(e^{-3x_i}+X^{\Lambda {\rm CDM} }e^{-4x_i} \right) +\Omega^{\Lambda {\rm CDM}}_{\Lambda} ,\\
\label{condiniexpR}
\mathcal{R}(x_i) & = 2+\frac{\Omega^{\Lambda {\rm CDM}}_{m}}{2\Omega^{\Lambda {\rm CDM}}_{\Lambda}}e^{-3x_i} .
\end{align}
\end{subequations}

In order to determine $z_i$, we assume that $f(R(z_i)) \simeq R - 2 \Lambda$. This condition can be expressed as follows \cite{Odintsov2017} :

\begin{equation}
e^{-\frac{R(z_i)}{\Lambda b}} \simeq \epsilon = 10^{-10} .
\end{equation}
In turn, this implies:
\begin{equation}
z_i = \left[\frac{ \Omega_\Lambda b}{\Omega_m }\left(\ln{\epsilon^{-1}-\dfrac{4}{b}}\right)\right]^{1/3}-1 .
\end{equation}
Thus, when $z>z_i$ we consider the solution of  the exponential model as the $\Lambda$CDM one 
and when $z<z_i$ the prediction of the model is calculated from the numerical integration of Eqs. $\eqref{expE}$ and $\eqref{expR}$. 

\subsection{Hu-Sawicki model}
\label{sec:HS}
For this model, the numerical integration of $H (z)$  performed with the change of variables proposed in \cite{Odintsov2017} is much more computationally expensive than the one accomplished with the proposal of de la Cruz-Dombriz \textit{et al.} \cite{2016PhRvD..93h4016D}.\footnote{Besides, the system of equations proposed in \cite{2016PhRvD..93h4016D} is also not the most appropriate for the exponential model.} Consequently, we implement the latter such that
 \begin{subequations}
\begin{align}
x & =\dfrac{\dot{R}f_{RR}}{Hf_{R}}\\
y & =\dfrac{f}{6H^{2}f_{R}}\\
v & =\dfrac{R}{6H^{2}}\\
\Omega & =\dfrac{8\pi G\rho_{m}}{3H^{2}f_{R}}\\
\Gamma & =\frac{f_{R}}{Rf_{RR}}\\
r & = R/R^{*},
\end{align}
\end{subequations}
where the constant $R^{*}$ has the same units as the Ricci scalar $R$ (in this
case, $R^{*} = R_{HS}$).
From this change of variables, the FLRW equations
(\ref{eq:Friedmanns}) and (\ref{eqrho}) become

\begin{subequations}
\begin{align}
\dfrac{dH}{dz} & = \frac{H}{z+1}\left(2-v\right)\\
\dfrac{dx}{dz} & =\dfrac{1}{z+1}\left(-\Omega-2v+x+4y+xv+x^{2}\right)\\
\dfrac{dy}{dz} & =\dfrac{-1}{z+1}\left(vx\Gamma-xy+4y-2yv\right)\\
\dfrac{dv}{dz} & =\dfrac{-v}{z+1}\left(x\Gamma+4-2v\right)\\
\dfrac{d\Omega}{dz} & =\dfrac{\Omega}{z+1}\left(-1+2v+x\right)\\
\dfrac{dr}{dz} & =-\dfrac{x\Gamma r}{z+1}.
\end{align}\label{eq:Sistema 2.0}
\end{subequations}

The latter system of equations is also solved numerically by defining the proper initial conditions.

When $b$ tends to zero, the numerical integration of Eqs. $\eqref{eq:Sistema 2.0}$  is particularly computationally expensive, becoming unstable for certain combinations of the parameters $b$ and $\Omega_m^0$. This occurs  because when the models ${\it f} (R)$ resemble $\Lambda$CDM,  $f_{RR}$ tends to zero. To avoid this problem, Basilakos \textit{et al.} \cite{PhysRevD.87.123529} proposed a method to obtain a series expansion of H(z)  around $b=0$ (the $\Lambda$CDM model solution).
In this way, there is no need to perform   the numerical integration in those regions of the parameter space that require large computational times. This approach was also used in many works such as \cite{2017JCAP...01..005N,PhysRevD.100.044041,PhysRevD.101.103505}. The general idea of this procedure is as follows; 
letting $N = -\log (1 + z)$ be the number of {\textit e}-foldings at redshift $z$, then
the Hubble parameter of the $\Lambda$CDM model can be written as
\begin{eqnarray}
H_{\Lambda {\rm CDM}}^{2}\left(N\right)&=&(H_{0}^{\Lambda {\rm CDM}})^{2}\Big[\Omega_{m}^{\Lambda {\rm CDM}}e^{-3N}\nonumber\\ &+&\left(1-\Omega_{m}^{\Lambda {\rm CDM}}\right)\Big],
\end{eqnarray}
and an expansion around it will be given by
\begin{equation}
 H^{2}\left(N\right)=H_{\Lambda {\rm CDM}}^{2}\left(N\right)+\sum_{i=1}^{M}b^{i}\delta H_{i}^{2}\left(N\right)\text{,}\label{eq:H_taylor_general}
\end{equation}
where $M$ is the number of terms that are used for the expansion. It has been studied  that, for the Hu-Sawicki model with $n = 1$,  the error in assuming expression \eqref{eq:H_taylor_general} just keeping the first two nonzero terms of the expansion (instead of the numerical integration) is of order  of $0.001\%$ for all redshifts  and  $b\leq 0.5$  (for details, see \cite{PhysRevD.87.123529}).
 Unfortunately, this method cannot be applied to the exponential ${\it f} (R)$ model since it cannot be expanded in series around $b = 0$.
 
 In a nutshell, for $b\leq 0.15$, we use Eq. \eqref{eq:H_taylor_general} up to order 2 in $b$, while for other values of $b$ we solve Eqs. \eqref{eq:Sistema 2.0} numerically. 
For this last case, as we did for the exponential model,  the initial conditions of the system of equations \eqref{eq:Sistema 2.0} are established so that the behavior of the $f(R)$ model  matches the one of the $\Lambda$CDM model.

\begin{subequations}
\begin{align}
x_{i} & = 0\\
y_{i}  & = \dfrac{(R_{i}-2 \Lambda)}{6H_{i}^{2}}\\
v_{i} & = \dfrac{R_{i}}{6H_{i}^{2}}\\
\Omega_{i} & = 1 - v_{i} + x_{i} + y_{i}\\
r_{i} & = R_{i}/R_{HS},
\end{align}
\end{subequations}

where $R_{i}=R^{{\rm \Lambda CDM}}\left(z_{i}\right)$ and $H_{i}=H^{{\rm \Lambda CDM}}\left(z_{i}\right)$ are the values of the Ricci tensor and the Hubble parameter on the initial condition, respectively. In this paper the initial redshift for the Hu-Sawicki model is set  at $z_i=10$. For both models, we have checked that the obtained solutions of the Friedmann equations do not depend on the particular choice of the initial redshift provided $z_i$ is sufficiently large ($z_i \ge 5$).\footnote{In fact, the percentage difference between solutions where $5 \le z_i < 10$ and the one assumed in this paper  is less than $0.3\%$.}
\newpage
\bibliographystyle{apsrev}
\bibliography{testmog}
\end{document}